\documentclass[12pt,a4paper]{article}
\usepackage{amssymb}
\usepackage{amsfonts}
\usepackage{amsmath}
\usepackage{float}
\usepackage{latexsym}
\usepackage{cite}
\begin{document}

\title{\vspace{-1cm}\bf Some properties of small perturbations\\ against a stationary solution of\\ the nonlinear Schr\"{o}dinger equation}

\author{
Mikhail~N.~Smolyakov$^{a,b}$
\\
$^a${\small{\em Skobeltsyn Institute of Nuclear Physics, Lomonosov Moscow
State University,
}}\\
{\small{\em Moscow 119991, Russia}}\\
$^b${\small{\em
Institute for Nuclear Research of the Russian Academy
of Sciences,}}\\
{\small{\em 60th October Anniversary prospect 7a, Moscow 117312,
Russia
}}}

\date{}
\maketitle

\begin{abstract}
In this paper, classical small perturbations against a stationary solution of the nonlinear Schr\"{o}dinger equation with the general form of nonlinearity are examined. It is shown that in order to obtain correct (in particular, conserved over time) nonzero expressions for the basic integrals of motion of a perturbation even in the quadratic order in the expansion parameter, it is necessary to consider nonlinear equations of motion for the perturbations. It is also shown that, despite the nonlinearity of the perturbations, the additivity property is valid for the integrals of motion of different nonlinear modes forming the perturbation (at least up to the second order in the expansion parameter).
\end{abstract}

\section{Introduction}
The nonlinear Schr\"{o}dinger equation is widely discussed in the scientific literature. For example, one can recall the Gross--Pitaevskii equation \cite{Gross,Pitaevskii1} providing an approximate description of a nonideal Bose gas at low temperatures (see also reviews \cite{DGPS,Pitaevskii2} and books \cite{Pitaevskii3,Pethick}) and its application in cosmology and astrophysics \cite{astro1,astro2,astro3}; nonlinear optics \cite{AA}; and, of course, various soliton solutions (see reviews \cite{review1,review2,review3}) including dark solitons \cite{review4,Frantzeskakis} and vortex solitons \cite{Malomed}. For a detailed discussion of some modern results in studying various systems described by the nonlinear Schr\"{o}dinger equation, see also \cite{LENCS}.

Among the variety of solutions of the nonlinear Schr\"{o}dinger equation, there exists a class of solutions that are stationary. The most known example of such a solution is the stationary solutions of the already mentioned Gross--Pitaevskii equation \cite{Gross,Pitaevskii1} describing a ground state of the Bose gas. Stationary solutions may also describe solitons, a simple analytical example of such a soliton in the case of logarithmic nonlinearity can be found in \cite{BB}.

Usually, small perturbations against a background solution are considered when it is necessary to study classical stability of this background solution. For the classical stability analysis, it is sufficient to consider only the linear approximation for the perturbations searching for exponentially growing modes, which is the standard technique. For example, in the case of a simplest stationary soliton solution of the nonlinear Schr\"{o}dinger equation, i.e., localized solution such that it is spherically symmetric, falls off to zero rapidly enough and has no nodes (like the one of \cite{BB}), this technique leads to the well-known Vakhitov-Kolokolov stability criterion established in \cite{VK,Kolokolov1,Kolokolov2}.

Meanwhile, small perturbations can be interesting by themselves. For example, it is well-known that perturbations against a stationary solution of the Gross--Pitaevskii equation describe phonons. Analogously, perturbations against a stationary soliton solution of the nonlinear Schr\"{o}dinger equation may provide an important information about its classical and even quantum properties. As an example, one can recall the role of perturbations in quantization of the kink solution, for a detailed discussion of this problem see \cite{Rajaraman:1982is} and references therein.

In the present paper, a detailed analysis of perturbations against a stationary solution of the nonlinear Schr\"{o}dinger equation is performed. It is shown that in order to obtain nonzero values of the basic integrals of motion of the perturbations, it is not sufficient to consider perturbations only in the linear approximation. In fact, the use of the linear approximation allows one to obtain only the zero values of the integrals of motion. In order to demonstrate this effect, the basic integrals of motion are calculated for the {\em nonlinear} classical perturbations against a stationary solution of the nonlinear Schr\"{o}dinger equation. These integrals of motion, which are the particle number, energy and momentum, as well as the corresponding equations of motion for the perturbations, are considered up to the quadratic order in the perturbations. Then, using the standard methods of perturbation theory, the integrals of motion are calculated up to the quadratic order in the corresponding expansion parameter, including the nonlinear corrections. It turned out that the use of nonlinear equations of motion for the perturbations is {\em necessary} for obtaining correct (in particular, conserved over time) expressions for the integrals of motion even in the quadratic order in the corresponding expansion parameter (i.e., in the lowest order providing nonzero values of these integrals of motion). Surprisingly, it also turns out that the additivity property is valid for such integrals of motion of different nonlinear modes, forming the perturbation against a stationary background solution. Of course, the result is not exact in the sense that it is proven only up to the second order in the corresponding expansion parameter. However, even in this approximation the effect could be important for a correct quantization of such systems. Note that this situation differs from the case of the kink solution, for which the use of only the linear approximation is sufficient to obtain correct expressions of the basic integrals of motion in the quadratic order in the corresponding expansion parameter \cite{Rajaraman:1982is}.

The effect described above can be considered from another point of view. Usually, the additivity property is valid for the integrals of motion of different dynamical systems if these systems do not interact with each other, which follows from the additivity of their Lagrangians \cite{LL}. Apart from this obvious case, the additivity property in a dynamical system is also connected with linearity of the corresponding equations of motion, as a consequence of the linear superposition principle. Although nothing forbids the emergence of such an additivity in nonlinear systems, such cases seem to be rather rare. As an explicit example of the additivity of energy in a nonlinear system, though in a somewhat specific form, one can recall the already mentioned paper \cite{BB}, in which the nonlinear Schr\"{o}dinger equation with logarithmic nonlinearity was examined. Thus, even though the additivity of the integrals of motion of different nonlinear modes, which is discussed in the present paper, is proven only up to the second order in the expansion parameter, but still it can be considered as an additional example of the additivity property in a nonlinear dynamical system. The latter can be interesting from a pure theoretical point of view -- in principle, this additivity property may indicate some sort of the nonlinear superposition principle.

One may expect that analogous effects exist in other nonlinear theories providing time-dependent background solutions. However, the use of the nonlinear Schr\"{o}dinger equation and its stationary solution as a background solution allows one to demonstrate the effect in a rather simple and explicit way.

\section{Setup and equations of motion for the perturbations}
Let us consider the nonlinear Schr\"{o}dinger equation in dimensionless variables
\begin{equation}\label{NSE}
i\frac{\partial\Psi}{\partial t}=-\Delta\Psi+V(\vec x)\Psi+F(\Psi^{*}\Psi)\Psi,
\end{equation}
where $V(\vec x)$ is the external potential, $\Delta=\sum\limits_{l=1}^{d}\partial_{l}^{2}$ and $d\ge 1$. It is well known that one can define the integrals of motion for this system, which are conserved over time if equation \eqref{NSE} holds. In particular, these are the particle number (norm)
\begin{equation*}
N=\int\Psi^{*}\Psi\,d^{d}x
\end{equation*}
(although the system is supposed to be classical, in what follows I will use the standard term ``particle number'') and the energy (Hamiltonian) of the system
\begin{equation*}
E=\int\left(\sum\limits_{l=1}^{d}\partial_{l}\Psi^{*}\partial_{l}\Psi+V\Psi^{*}\Psi+\int\limits_{0}^{\Psi^{*}\Psi}F(s)ds\right)d^{d}x.
\end{equation*}
If $V(\vec x)\equiv 0$, the momentum defined as
\begin{equation*}
P_{l}=\frac{i}{2}\int \left(\Psi\partial_{l}\Psi^{*}-\Psi^{*}\partial_{l}\Psi\right)d^{d}x,
\end{equation*}
where $l=1,...,d$, is also conserved over time. It is clear that system \eqref{NSE} possess the global $U(1)$ symmetry.

Suppose we have a stationary solution to equation \eqref{NSE} of the form
\begin{equation}\label{backgrsol}
\Psi_{0}(t,\vec x)=e^{-i\omega t}f(\vec x),
\end{equation}
where the function $f(\vec x)$ is real. In order to simplify the subsequent analysis and ensure the finiteness of the particle number and energy of the stationary solution (for example, if $f(\vec x)\equiv\textrm{const}\neq 0$), one can put the system into a ``box'' of a finite size if necessary.

Now we consider a small perturbation against this background solution such that
\begin{equation}\label{psipert}
\Psi(t,\vec x)=e^{-i\omega t}\left(f(\vec x)+\alpha\varphi(t,\vec x)\right).
\end{equation}
Here $\alpha\ll 1$ is real and $\varphi(t,\vec x)$ is supposed to be of the order of $f(\vec x)$. As was noted in the Introduction, usually the perturbations against stationary solutions of the nonlinear Schr\"{o}dinger equation are considered in the linear approximation for the perturbations, like in the case of the linear stability analysis \cite{VK,Kolokolov1,Kolokolov2}. Contrary to this standard approach, let us obtain the equation of motion for the perturbation $\varphi$ up to the quadratic order in $\varphi$. The corresponding equation of motion takes the form
\begin{equation}\label{eqpert}
i\frac{\partial\varphi}{\partial t}=-\Delta\varphi+(V+U-\omega)\varphi+fW(\varphi+\varphi^{*})+\alpha\left(W(\varphi^{2}+2\varphi^{*}\varphi)+J(\varphi+\varphi^{*})^{2}\right),
\end{equation}
where
\begin{equation*}
U(\vec x)=F\left(f^{2}(\vec x)\right),\qquad W(\vec x)=\frac{dF(s)}{ds}\biggl|_{s=f^{2}(\vec x)}f(\vec x),\qquad J(\vec x)=\frac{1}{2}\frac{d^{2}F(s)}{ds^{2}}\biggl|_{s=f^{2}(\vec x)}f^{3}(\vec x).
\end{equation*}
A solution to equation \eqref{eqpert} can be represented as a series in the small expansion parameter $\alpha$. However, since equation \eqref{eqpert} is valid up to the linear order in $\alpha$ (the terms of the higher orders are omitted), it does not make sense to consider terms of the second order in $\alpha$ and higher in the perturbation. Thus, I take the following ansatz for the perturbation, which is suggested by the form of equation \eqref{eqpert} (to simplify the analysis, below I will consider only the oscillation modes):
\begin{align}\nonumber
2\textrm{Re}(\varphi)=\sum_{n}\left(\xi_{n}e^{-i\gamma_{n}t}+\xi^{*}_{n}e^{i\gamma_{n}t}\right)+\alpha\chi_{+}+
\alpha\sum_{n}\left(\psi_{+,n}e^{-2i\gamma_{n}t}+\psi_{+,n}^{*}e^{2i\gamma_{n}t}\right)\\\label{substosc1}
+\alpha\sum_{n,k;\,n<k}\left(\varrho_{+,nk}e^{-i(\gamma_{n}+\gamma_{k})t}+\varrho_{+,nk}^{*}e^{i(\gamma_{n}+\gamma_{k})t}+
\theta_{+,nk}e^{-i(\gamma_{n}-\gamma_{k})t}+\theta_{+,nk}^{*}e^{i(\gamma_{n}-\gamma_{k})t}\right),\\\nonumber
2i\textrm{Im}(\varphi)=\sum_{n}\left(\eta_{n}e^{-i\gamma_{n}t}-\eta^{*}_{n}e^{i\gamma_{n}t}\right)+\alpha\chi_{-}+
\alpha\sum_{n}\left(\psi_{-,n}e^{-2i\gamma_{n}t}-\psi_{-,n}^{*}e^{2i\gamma_{n}t}\right)\\\label{substosc2}
+\alpha\sum_{n,k;\,n<k}\left(\varrho_{-,nk}e^{-i(\gamma_{n}+\gamma_{k})t}-\varrho_{-,nk}^{*}e^{i(\gamma_{n}+\gamma_{k})t}+
\theta_{-,nk}e^{-i(\gamma_{n}-\gamma_{k})t}-\theta_{-,nk}^{*}e^{i(\gamma_{n}-\gamma_{k})t}\right).
\end{align}
Here $\xi_{n}(\vec x)$, $\eta_{n}(\vec x)$, $\chi_{-}(\vec x)$, $\psi_{\pm,n}(\vec x)$, $\varrho_{\pm,nk}(\vec x)$, $\theta_{\pm,nk}(\vec x)$ are complex functions, $\chi_{+}(\vec x)$ is a real function. At this step I also suppose that $\gamma_{n}\neq\gamma_{k}$ for $n\neq k$. One can see that the terms of the zero order in $\alpha$ represent the standard form of a perturbation composed from the oscillation modes in the case of the linear approximation \cite{DGPS,AA,Frantzeskakis}. Since $\gamma_{n}$ is real for oscillation modes \cite{AA}, without loss of generality one can set $\gamma_{n}>0$. For simplicity, here I suppose that the spectrum of the modes is discrete. However, the modes from the continuous spectrum, if exist, can be easily taken into account: the simplest way to do it is to put the system into a ``box'' of a finite size.

Note that the linearized equation of motion for perturbations also provides modes which have completely different forms, i.e., nonoscillation modes. Apart from the trivial cases of translational modes $\partial_{l}f$ and the mode $if$, there may exist the exponentially growing instability mode\footnote{Since the exponentially growing instability mode destroys the corresponding background solution, practically it does not make sense to examine perturbations against such a background in detail. In order not to deal with exponentially growing modes, one can consider only classically stable background solutions. As was noted in the Introduction, if $f(\vec x)$ is spherically symmetric, falls off to zero rapidly enough and has no nodes, for $V(\vec x)\equiv 0$ such background solutions can be selected using the Vakhitov-Kolokolov stability criterion established in \cite{VK,Kolokolov1,Kolokolov2}. Soliton-like ``bubbles'' of form \eqref{backgrsol} for $V(\vec x)\equiv 0$ are always classically unstable \cite{Barashenkov1,Barashenkov2,Debouard}. Meanwhile, there may exist classically stable dark solitons for some $V(\vec x)\not\equiv 0$ \cite{Frantzeskakis}.}, the mode corresponding to Galilean transformations and the mode $\frac{df}{d\omega}-itf$,
which corresponds to the change of the frequency $\omega$ of background solution \eqref{backgrsol}. Below I will not examine nonlinear corrections produced by these modes and their nonlinear interaction with oscillation modes, such an analysis calls for additional detailed investigation.

Substituting \eqref{substosc1} and \eqref{substosc2} into equation \eqref{eqpert} and keeping the terms up to the linear order in $\alpha$, one can get the following set of equations:
\begin{align}\label{eqxin}
&\hat L_{1}\xi_{n}=\gamma_{n}\eta_{n},\\\label{eqetan}
&\hat L_{2}\eta_{n}=\gamma_{n}\xi_{n},\\
\label{chipluseq}
&\hat L_{1}\chi_{+,n}=-W\left(3\xi_{n}^{*}\xi_{n}+\eta_{n}^{*}\eta_{n}\right)-4J\xi_{n}^{*}\xi_{n},\\
\label{chiminuseq}
&\hat L_{2}\chi_{-,n}=-W\left(\xi_{n}^{*}\eta_{n}-\eta_{n}^{*}\xi_{n}\right),\\
\label{psipluseq}
&\hat L_{1}\psi_{+,n}=2\gamma_{n}\psi_{-,n}-W\left(\frac{3}{2}\xi_{n}^{2}-\frac{1}{2}\eta_{n}^{2}\right)-2J\xi_{n}^{2},\\\label{psiminuseq}
&\hat L_{2}\psi_{-,n}=2\gamma_{n}\psi_{+,n}-W\xi_{n}\eta_{n},\\
\label{eqvarrhoplus}
&\hat L_{1}\varrho_{+,nk}=(\gamma_{n}+\gamma_{k})\varrho_{-,nk}-W\left(3\xi_{n}\xi_{k}-\eta_{n}\eta_{k}\right)-4J\xi_{n}\xi_{k},\\\label{eqvarrho}
&\hat L_{2}\varrho_{-,nk}=(\gamma_{n}+\gamma_{k})\varrho_{+,nk}-W\left(\xi_{n}\eta_{k}+\eta_{n}\xi_{k}\right),\\
\label{eqthetaplus}
&\hat L_{1}\theta_{+,nk}=(\gamma_{n}-\gamma_{k})\theta_{-,nk}-W\left(3\xi_{n}\xi_{k}^{*}+\eta_{n}\eta_{k}^{*}\right)-4J\xi_{n}\xi_{k}^{*},\\
\label{eqtheta}
&\hat L_{2}\theta_{-,nk}=(\gamma_{n}-\gamma_{k})\theta_{+,nk}-W\left(\eta_{n}\xi_{k}^{*}-\xi_{n}\eta_{k}^{*}\right),
\end{align}
where $\chi_{+}=\sum\limits_{n}\chi_{+,n}$, $\chi_{-}=\sum\limits_{n}\chi_{-,n}$ and
\begin{equation}\label{Leq}
\hat L_{2}=-\Delta+V+U-\omega,\qquad \hat L_{1}=\hat L_{2}+2fW.
\end{equation}
As expected, equations \eqref{eqxin} and \eqref{eqetan} are the standard equations of motion for oscillation modes in the linear approximation.

It is clear that since the perturbation $\varphi$ in \eqref{substosc1} and \eqref{substosc2} is a solution of nonlinear equation of motion \eqref{eqpert}, it is impossible to represent $\varphi$ as a sum of different modes (like in the case of linear approximation). However, it is convenient to define the ``nonlinear mode'' as the part of the nonlinear perturbation $\varphi$ which is characterized by the frequency $\gamma_{n}$, the functions $\xi_{n}$, $\eta_{n}$, and the functions $\chi_{\pm,n}$, $\psi_{\pm,n}$. All the other terms in $\varphi$ (characterized by the functions $\varrho_{\pm,nk}$ and $\theta_{\pm,nk}$) describe the overlap terms between different nonlinear modes due to the nonlinearity of the theory.

\section{Integrals of motion of the perturbations}
The particle number, the energy and the momentum of the perturbation are defined in the standard way as
\begin{equation*}
N_{p}=N[\Psi(t,\vec x)]-N[\Psi_{0}(t,\vec x)],\quad
E_{p}=E[\Psi(t,\vec x)]-E[\Psi_{0}(t,\vec x)],\quad P_{p,l}=P_{l}[\Psi(t,\vec x)],
\end{equation*}
where $\Psi(t,\vec x)$ is defined by \eqref{psipert}. Expanding $E[\Psi(t,\vec x)]$ up to the second order in $\alpha$ (again, since equation \eqref{eqpert} is valid up to the linear order in $\alpha$, it does not make sense to consider terms of the higher orders in $\alpha$) and using equation of motion \eqref{eqpert} when deriving $E_{p}$, after some algebra one can get (the expressions for $N_{p}$ and $P_{p,l}$ are exact)
\begin{align}\label{Npert}
&N_{p}=\alpha\int d^{d}x\left(f(\varphi+\varphi^{*})+\alpha\varphi^{*}\varphi\right),\\\label{Epert}
&E_{p}=\omega N_{p}+\frac{i\alpha^{2}}{2}\int d^{d}x\left(\varphi^{*}\frac{\partial\varphi}{\partial t}-\frac{\partial\varphi^{*}}{\partial t}\varphi\right),\\
&P_{p,l}=i\alpha\int\left(\partial_{l}f(\varphi-\varphi^{*})-\alpha\varphi^{*}\partial_{l}\varphi\right)d^{d}x.
\end{align}

\subsection{Particle number}
Let us substitute \eqref{substosc1} and \eqref{substosc2} into \eqref{Npert} and keep the terms up to the second order in $\alpha$. The result looks as follows
\begin{align}\nonumber
&N_{p}=\alpha\sum_{n}\left[e^{-i\gamma_{n}t}\int f\xi_{n}d^{d}x+\textrm{c.c.}\right]+\alpha^{2}\int\left(f\chi_{+}+\frac{1}{2}\sum_{n}\left(\xi_{n}\xi_{n}^{*}+\eta_{n}\eta_{n}^{*}\right)\right)d^{d}x
\\\nonumber&+
\alpha^{2}\sum_{n}\left[e^{-2i\gamma_{n}t}\int\left(f\psi_{+,n}+\frac{1}{4}(\xi_{n}^{2}-\eta_{n}^{2})\right)d^{d}x+\textrm{c.c.}\right]\\
\nonumber&+
\alpha^{2} \sum_{n,k;\,n<k}\left[e^{-i(\gamma_{n}+\gamma_{k})t}\int\left(f\varrho_{+,nk}+\frac{1}{2}(\xi_{n}\xi_{k}-\eta_{n}\eta_{k})\right)d^{d}x+\textrm{c.c.}\right]
\\\label{pnumber} &+\alpha^{2}\sum_{n,k;\,n<k}\left[e^{-i(\gamma_{n}-\gamma_{k})t}\int\left(f\theta_{+,nk}+\frac{1}{2}(\xi_{n}\xi_{k}^{*}+\eta_{n}\eta_{k}^{*})\right) d^{d}x+\textrm{c.c.}\right].
\end{align}
It turns out that all terms in the square brackets, i.e., the terms, which explicitly depend on time, vanish. For the term linear in $\alpha$ this can be shown by multiplying equation \eqref{eqetan} by $f$, integrating over the space and using the facts that $\gamma_{n}\neq 0$ and $\hat L_{2}f=0$ (the latter equation is just equation \eqref{NSE} for background solution \eqref{backgrsol}). As for the other terms, the detailed calculations can be found in Appendix~A. Thus, for the particle number of perturbation \eqref{substosc1}, \eqref{substosc2} we get
\begin{equation}\label{chargesuper}
N_{p}=\alpha^{2}\sum_{n}\int\left(f\chi_{+,n}+\frac{1}{2}\left(\xi_{n}\xi_{n}^{*}+\eta_{n}\eta_{n}^{*}\right)\right)d^{d}x,
\end{equation}
which is just the sum of the particle numbers $N_{p}^{n}$ of each nonlinear mode. The first term in the brackets of \eqref{chargesuper} originates from the nonlinear ($\sim\alpha$) part of the solution for the perturbation $\varphi$, but comes from the terms of \eqref{Npert} that are linear in $\varphi$ and $\varphi^{*}$; whereas the other terms in the brackets of \eqref{chargesuper} originate from the linear part of the solution for the perturbation $\varphi$, but come from the term of \eqref{Npert} that is quadratic in $|\varphi|$.

An important part of the derivation is the assumption $\gamma_{n}\neq\gamma_{k}$ for $n\neq k$. But what if we have two modes such that $\gamma_{n}=\gamma_{k}=\gamma$ for $\xi_{n}\not\equiv c\xi_{k}$, $\eta_{n}\not\equiv c\eta_{k}$, where $c$ is a constant, i.e., different modes of the same frequency? In this case one gets in \eqref{chargesuper} $\xi_{n}+\xi_{k}$ instead of a single $\xi$ and $\eta_{n}+\eta_{k}$ instead of a single $\eta$. There is a good reason to believe that the corresponding overlap integrals in \eqref{chargesuper} are equal to zero, i.e.,
\begin{equation*}
N_{p}^{n+k}[\xi_{n}+\xi_{k},\eta_{n}+\eta_{k},\gamma]=N_{p}^{n}[\xi_{n},\eta_{n},\gamma]+N_{p}^{k}[\xi_{k},\eta_{k},\gamma].
\end{equation*}
In order to show it, I will use the standard trick and modify ``by hands'' the function $W(\vec x)\to W_{(\epsilon)}(\vec x)=W(\vec x)+\epsilon\,\delta W(\vec x)$, where $\delta W(\vec x)$ is chosen in such a way that it removes the degeneracy of the modes
\begin{equation*}
\gamma_{n}\to \gamma_{n}^{(\epsilon)}=\gamma+\epsilon\,\delta\gamma_{n},\qquad \gamma_{k}\to \gamma_{k}^{(\epsilon)}=\gamma+\epsilon\,\delta\gamma_{k},
\end{equation*}
where $\delta\gamma_{n}\neq \delta\gamma_{k}$. Of course, the functions $\xi_{n}$, $\xi_{k}$ and $\eta_{n}$, $\eta_{k}$ (as well as the functions $\chi_{+,n}$, $\chi_{+,k}$, which are expressed through $\xi_{n}$, $\eta_{n}$ and $\xi_{k}$, $\eta_{k}$ by means of equation \eqref{chipluseq}) also turn out to be modified as $\xi_{n}, \xi_{k}\to \xi_{n}^{(\epsilon)}, \xi_{k}^{(\epsilon)}$ and $\eta_{n}, \eta_{k}\to \eta_{n}^{(\epsilon)}, \eta_{k}^{(\epsilon)}$. In this case $\gamma_{n}^{(\epsilon)}\neq\gamma_{k}^{(\epsilon)}$ and, according to the results presented above (the modification of the function $W$ does not change the operator $\hat L_{2}$, so the necessary equation $\hat L_{2}f=0$ remains intact, see Appendix A), the corresponding overlap terms vanish, so we are left with
\begin{equation*}
N_{p}^{n}[\xi_{n}^{(\epsilon)},\eta_{n}^{(\epsilon)},\gamma_{n}^{(\epsilon)}]+N_{p}^{k}[\xi_{k}^{(\epsilon)},\eta_{k}^{(\epsilon)},\gamma_{k}^{(\epsilon)}].
\end{equation*}
In the limit $\epsilon\to 0$ we still get
\begin{equation*}
N_{p}^{n}[\xi_{n},\eta_{n},\gamma]+N_{p}^{k}[\xi_{k},\eta_{k},\gamma]
\end{equation*}
without any cross terms. Analogous reasonings can be applied to the cases of energy and momentum (with some modifications for the latter), which will be discussed below.

It is not necessary to solve equation \eqref{chipluseq} for each $n$. If we multiply equation \eqref{chipluseq} by $\frac{df}{d\omega}$, integrate the result over the space and use the fact that $\hat L_{1}\frac{df}{d\omega}=f$ (which can be obtained by differentiating the equation $\hat L_{2}f=0$ with respect to $\omega$), we will obtain for \eqref{chargesuper}
\begin{equation}\label{PNfinal}
N_{p}=\alpha^{2}\sum_{n}\int\left(\frac{1}{2}\left(\xi_{n}^{*}\xi_{n}+\eta_{n}^{*}\eta_{n}\right)-\frac{df}{d\omega}W\left(3\xi_{n}^{*}\xi_{n}+\eta_{n}^{*}\eta_{n}\right)
-4\frac{df}{d\omega}J\xi_{n}^{*}\xi_{n}\right)d^{d}x,
\end{equation}
which formally depends only on $\xi_{n}$ and $\eta_{n}$. Note that the sign of $N_{p}^{n}$ is not definite in the general case. The existence of the term with the function $J$ in \eqref{PNfinal} (the function $J$ comes from the nonlinear term in equation of motion \eqref{eqpert}) is an additional demonstration of the fact that contribution of the nonlinear corrections is indeed nonzero.

\subsection{Energy}
Now I turn to the energy of perturbation \eqref{substosc1}, \eqref{substosc2}. Substituting it into \eqref{Epert} and keeping the terms up to the second order in $\alpha$, one gets
\begin{align}\nonumber
&E_{p}=\omega N_{p}+\frac{\alpha^{2}}{2}\sum_{n}\left(\gamma_{n}\int \left(\xi_{n}\eta_{n}^{*}+\xi_{n}^{*}\eta_{n}\right)d^{d}x\right)\\\nonumber&+\frac{\alpha^{2}}{4}\sum_{n,k;\,n<k}\left[
e^{-i(\gamma_{n}+\gamma_{k})t}(\gamma_{n}-\gamma_{k})\int\left(\eta_{n}\xi_{k}-\eta_{k}\xi_{n}\right)d^{d}x+\textrm{c.c.}\right]
\\\label{energyprelim}&
+\frac{\alpha^{2}}{4}\sum_{n,k;\,n<k}\left[
e^{-i(\gamma_{n}-\gamma_{k})t}(\gamma_{n}+\gamma_{k})\int\left(\eta_{n}\xi_{k}^{*}+\eta_{k}^{*}\xi_{n}\right)d^{d}x+\textrm{c.c.}
\right].
\end{align}
Again, all the time-dependent terms in \eqref{energyprelim} vanish, see Appendix~B for details. Thus, one obtains
\begin{equation}\label{Efinal}
E_{p}=\sum_{n}\left(
\omega N_{p}^{n}+\frac{\alpha^{2}}{2}\gamma_{n}\int\left(\xi_{n}\eta_{n}^{*}+\xi_{n}^{*}\eta_{n}\right)d^{d}x\right),
\end{equation}
which is also just the sum of the energies $E_{p}^{n}$ of each nonlinear mode.

It is interesting to note that the sign of $\gamma_{n}\int\left(\xi_{n}\eta_{n}^{*}+\xi_{n}^{*}\eta_{n}\right)d^{d}x$ is just the Krein signature of the mode \cite{Frantzeskakis}. Usually, the Krein signatures of the modes are positive. However, in the case of dark solitons there may exist an anomalous mode such that its Krein signature is negative \cite{Frantzeskakis}.

Although the particle number of each mode is nonzero in the general case, it is possible to create a nonlinear perturbation in the processes which do not change the total particle number of the system. Indeed, such a process may look like modification of the initial background solution and creation of the perturbation against this new background solution such that
\begin{equation}\label{deltaN}
N[\Psi_{0}^{\omega+\Delta\omega}(t,\vec x)]+N_{p}(\omega+\Delta\omega)=N[\Psi_{0}^{\omega}(t,\vec x)].
\end{equation}
Since $N_{p}\sim\alpha^{2}$, the change of the frequency $\omega$ of the background solution is such that $\Delta\omega\sim\alpha^{2}$, so we can use $N_{p}(\omega+\Delta\omega)\approx N_{p}(\omega)$ (as well as $E_{p}(\omega+\Delta\omega)\approx E_{p}(\omega)$) with the same accuracy. Then, for the total energy we can write
\begin{align*}
E[\Psi_{0}^{\omega+\Delta\omega}(t,\vec x)]+E_{p}(\omega+\Delta\omega)\approx E[\Psi_{0}^{\omega}(t,\vec x)]+\Delta\omega\frac{dE[\Psi_{0}^{\omega}(t,\vec x)]}{d\omega}+E_{p}(\omega)\\
=E[\Psi_{0}^{\omega}(t,\vec x)]+\Delta\omega\,\omega\,\frac{dN[\Psi_{0}^{\omega}(t,\vec x)]}{d\omega}+E_{p}(\omega)\approx
E[\Psi_{0}^{\omega}(t,\vec x)]+\omega\Delta N+E_{p}(\omega),
\end{align*}
where we have used the relation \cite{review1}
\begin{equation*}
\frac{dE[\Psi_{0}^{\omega}(t,\vec x)]}{d\omega}=\omega\frac{dN[\Psi_{0}^{\omega}(t,\vec x)]}{d\omega},
\end{equation*}
which holds for any background configuration \eqref{backgrsol}.
According to \eqref{deltaN}, $\Delta N\approx -N_{p}(\omega)$. Thus, using \eqref{Efinal}, we obtain for the total energy of the final configuration
\begin{equation}\label{Etotfinal}
E[\Psi_{0}^{\omega}(t,\vec x)]+\frac{\alpha^{2}}{2}\sum_{n}\left(\gamma_{n}\int\left(\xi_{n}\eta_{n}^{*}+\xi_{n}^{*}\eta_{n}\right)d^{d}x\right).
\end{equation}
That is, in the processes conserving the particle number the part $\omega N_{p}$ of the perturbation energy $E_{p}$ is connected with the background solution: if $\omega N_{p}>0$, then the energy $\omega N_{p}$ is released by the background solution; if $\omega N_{p}<0$, then the energy $|\omega N_{p}|$ is absorbed by the background solution.

It is worth mentioning that the modification of the initial background solution (i.e., the change of the frequency $\omega$ of the background solution described above) is equivalent to creation of the nonoscillation mode $\frac{df}{d\omega}-itf$. Indeed, let us add the term
\begin{equation}\label{nonoscomega}
\frac{1}{\alpha}\Delta\omega\left(\frac{df}{d\omega}-itf\right)
\end{equation}
with $\Delta\omega=-\left(\frac{dN[\Psi_{0}^{\omega}(t,\vec x)]}{d\omega}\right)^{-1}N_{p}(\omega)$ to the perturbation $\varphi$ consisting of oscillation modes. Substituting \eqref{nonoscomega} into \eqref{Epert} with \eqref{Npert} and using the fact that $\Delta\omega\sim\alpha^{2}$, for the whole perturbation we get up to the terms $\sim\alpha^{2}$ exactly \eqref{Etotfinal} .

\subsection{Momentum}
As for the momentum of the perturbation, we get
\begin{align}\nonumber
&P_{p,l}=i\alpha\sum_{n}\left[e^{-i\gamma_{n}t}\int\partial_{l}f\eta_{n}d^{d}x-\textrm{c.c.}\right]+
i\alpha^{2}\int \left(\partial_{l}f\chi_{-}-\frac{1}{2}\sum_{n}\Bigl(\xi_{n}^{*}\partial_{l}\eta_{n}-\xi_{n}\partial_{l}\eta_{n}^{*}\Bigr)\right)d^{d}x\\\nonumber&+
i\alpha^{2}\sum_{n}\left[e^{-2i\gamma_{n}}\int\left(\partial_{l}f\psi_{-,n}-\frac{1}{2}\xi_{n}\partial_{l}\eta_{n}\right)d^{d}x-\textrm{c.c.}\right]
\\
\nonumber&+
i\alpha^{2} \sum_{n,k;\,n<k}\left[e^{-i(\gamma_{n}+\gamma_{k})t}\int\left(\partial_{l}f\varrho_{-,nk}-\frac{1}{2}\left(\xi_{k}\partial_{l}\eta_{n}+
\xi_{n}\partial_{l}\eta_{k}\right)\right)d^{d}x
-\textrm{c.c.}\right]\\\label{momentum}&+
i\alpha^{2}\sum_{n,k;\,n<k}
\left[e^{-i(\gamma_{n}-\gamma_{k})t}\int\left(\partial_{l}f\theta_{-,nk}-\frac{1}{2}\left(\xi_{k}^{*}\partial_{l}\eta_{n}-
\xi_{n}\partial_{l}\eta_{k}^{*}\right)\right)d^{d}x-\textrm{c.c.}\right].
\end{align}
Let $V(\vec x)\equiv 0$. Then, all the time-dependent terms in the latter formula (i.e., the terms in the square brackets) vanish. For the term linear in $\alpha$ this can be shown by multiplying equation \eqref{eqxin} by $\partial_{l}f$, integrating over the space and using the facts that $\gamma_{n}\neq 0$ and $\hat L_{1}\partial_{l}f=0$ if $V(\vec x)\equiv 0$ (the latter equation can be obtained by differentiating the equation $\hat L_{2}f=0$ with $V(\vec x)\equiv 0$ with respect to $x^{l}$). As for the other time-dependent terms, the detailed calculations can be found in Appendix~C. Thus, one obtains\footnote{In order to deal with the modes such that $\gamma_{n}=\gamma_{k}=\gamma$ for $\xi_{n}\not\equiv c\xi_{k}$, $\eta_{n}\not\equiv c\eta_{k}$, where $c$ is a constant (see the discussion in Subsection~3.1), one should modify ``by hands'' the functions $W(\vec x)\to W_{(\epsilon)}(\vec x)=W(\vec x)+\epsilon\,\delta W(\vec x)$ and $U(\vec x)\to U_{(\epsilon)}(\vec x)=U(\vec x)+\epsilon\,\delta U(\vec x)$ in such a way that $\delta U\equiv-2f\delta W$. In this case, the necessary equation $\hat L_{1}\partial_{l}f=0$ remains intact, see Appendix C.}
\begin{equation}\label{momentumsuper}
P_{p,l}=i\alpha^{2}\sum_{n}\int \left(\partial_{l}f\chi_{-,n}-\frac{1}{2}\Bigl(\xi_{n}^{*}\partial_{l}\eta_{n}-\xi_{n}\partial_{l}\eta_{n}^{*}\Bigr)\right)d^{d}x.
\end{equation}
Again, the total momentum of the perturbation is the sum of the momenta of each nonlinear mode.

As in the case of the particle number, it is not necessary to solve equation \eqref{chiminuseq} for each $n$. If $f(\vec x)\not\equiv\textrm{const}$, for the given $f(\vec x)$ one can solve once the equation
\begin{equation}\label{auxg}
\hat L_{2}g_{l}=\partial_{l}f.
\end{equation}
Then, multiplying equation \eqref{chiminuseq} by $g_{l}(\vec x)$, integrating the result over the space and using equation \eqref{auxg}, we get for \eqref{momentumsuper}
\begin{equation}\label{Pfinal}
P_{p,l}=-i\alpha^{2}\sum_{n}\int \left(\frac{1}{2}\left(\xi_{n}^{*}\partial_{l}\eta_{n}-\xi_{n}\partial_{l}\eta_{n}^{*}\right)
+g_{l}W\left(\xi_{n}^{*}\eta_{n}-\eta_{n}^{*}\xi_{n}\right)\right)d^{d}x,
\end{equation}
which formally depends only on $\xi_{n}$ and $\eta_{n}$.

\subsection{Small discussion}
One can see that though  perturbation \eqref{substosc1}, \eqref{substosc2} satisfies the set of equations \eqref{eqxin}--\eqref{eqtheta}, which follows from {\em nonlinear} equation of motion \eqref{eqpert}, and contains explicit terms describing nonlinear corrections (the terms with $\chi_{\pm,n}$ and $\psi_{\pm,n}$) and overlapping between different modes (the terms with $\varrho_{\pm,nk}$ and $\theta_{\pm,nk}$), the resulting expressions for the particle number, energy and momentum of the perturbation do not contain any terms describing interaction between different modes. That is, the total particle number, energy and momentum of the nonlinear perturbation, at least up to the quadratic order in the expansion parameter $\alpha$, are just the {\em exact} sums of the corresponding expressions for each of the nonlinear mode itself, which is nothing but a manifestation of the additivity property. In fact, it means that there is no physical interaction between different nonlinear modes up to the quadratic order in the expansion parameter $\alpha$.

However, the terms with $\psi_{\pm,n}$, $\varrho_{\pm,nk}$ and $\theta_{\pm,nk}$ in \eqref{substosc1}, \eqref{substosc2} are very important. Suppose we take just the linear approximation for the perturbation. In this case we will get formulas \eqref{pnumber} and \eqref{momentum} without the nonlinear corrections, and in the general case the time-dependent terms will not vanish. The latter implies not only the absence of the additivity effect described above, but also non-conservation over time of the particle number, energy and momentum, which will be explicitly demonstrated in the next section. The origin of this non-conservation is trivial. Indeed, the integrals of motion are conserved over time if the equation of motion is fulfilled; thus, if the equation of motion for the perturbations is linear, one can expect that the corresponding integrals of motion are conserved only in the linear order in the perturbations (in fact, for the oscillation modes they are equal to zero in this approximation, which is confirmed by the absence of the terms $\sim\alpha$ in \eqref{PNfinal}, \eqref{Efinal} and \eqref{Pfinal}) and it is incorrect to consider expressions which are quadratic in the perturbations. Thus, the use of the nonlinear equation of motion for the perturbation results not only in the additivity of the integrals of motion of different nonlinear modes, but also ensures the conservation over time of these integrals of motion and provides their correct (recall the contributions of $\chi_{\pm,n}$ in equations \eqref{chargesuper} and \eqref{momentumsuper}) nonzero values up to the quadratic order in the expansion parameter $\alpha$. The latter is important even without any reference to the additivity effect, but for obtaining the correct values of the basic physical characteristics of just a single oscillation mode.

Note that if $f(\vec x)\equiv\textrm{const}$ (the case of a condensate), then some of the time-dependent terms in formulas \eqref{pnumber} and \eqref{momentum} may vanish even if the equation of motion for the perturbations is linear (i.e., if there are no terms with $\chi_{\pm,n}$, $\psi_{\pm,n}$, $\varrho_{\pm,nk}$ and $\theta_{\pm,nk}$ in formulas \eqref{pnumber} and \eqref{momentum}). This happens when a perturbation consists of just the plane waves $\sim e^{\pm i(\gamma_{n}t-\vec k_{n}\vec x)}$. However, the contributions of the terms with $\chi_{+,n}$, which are of the same order as the other time-independent terms $\sim\alpha^{2}$, are missed in such a case. This example will be discussed in the next section.

\section{Explicit examples}
\subsection{Logarithmic nonlinearity}
In the present subsection, let us study the perturbations in the model with logarithmic nonlinearity and without external potential, which was proposed and examined in \cite{BB} (see also \cite{NLSlog1,NLSlog2} and \cite{Hefter} for applications of the nonlinear Schr\"{o}dinger equation with such a logarithmic nonlinearity in nonlinear optics and even in nuclear physics). The main aim of this example is to show that without the nonlinear corrections, the integrals of motion calculated up to the quadratic order in perturbations indeed are not conserved over time in the general case.

Let $V(\vec x)\equiv 0$ and
\begin{equation}\label{log-pot-F}
F(\Psi^{*}\Psi)=-\ln\left(\Psi^{*}\Psi\right).
\end{equation}
Stationary solution, corresponding to this form of nonlinearity, has the form
\begin{equation}\label{sol-pot-F}
\Psi(t,\vec x)=e^{-i\omega t}Ae^{-\frac{{\vec x}^{2}}{2}},\qquad \omega=d-\ln(A^{2}),
\end{equation}
where $A$ is a real constant. Equations \eqref{eqxin} and \eqref{eqetan} for the linear mode take the form
\begin{align*}
&\left(-\Delta+{\vec x}^{2}-2-d\right)\xi_{n}=\gamma_{n}\eta_{n},\\
&\left(-\Delta+{\vec x}^{2}-d\right)\eta_{n}=\gamma_{n}\xi_{n}.
\end{align*}
It is clear that solutions to these equations have the form
\begin{equation}\label{YZG}
\xi_{\hat n}(\vec x)=Y_{\hat n}G_{\hat n}(\vec x),\qquad \eta_{\hat n}(\vec x)=Z_{\hat n}G_{\hat n}(\vec x),
\end{equation}
with
\begin{equation*}
\left(-\Delta+{\vec x}^{2}\right)G_{\hat n}=\lambda_{\hat n}G_{\hat n},\qquad \lambda_{\hat n}=d+2(n_{1}+...+n_{d}),
\end{equation*}
where $\hat n=\{n_{1},...,n_{d}\}$, $n_{1,...,d}=0,1,2,...,d$, the subscript $\hat n$ is used instead of the subscript $n$, and the coefficients $Y_{\hat n}$, $Z_{\hat n}$ satisfy the system of equations
\begin{align}\label{YZ1}
&(\lambda_{\hat n}-2-d)Y_{\hat n}-\gamma_{\hat n}Z_{\hat n}=0,\\\label{YZ2}
&-\gamma_{\hat n}Y_{\hat n}+(\lambda_{\hat n}-d)Z_{\hat n}=0.
\end{align}
The functions $G_{\hat n}(\vec x)=G_{n_{1}}(x_{1})\times G_{n_{2}}(x_{2})\times...\times G_{n_{d}}(x_{d})$ in \eqref{YZG} can be taken to be real. For $n_{1}+...+n_{d}>1$, equations \eqref{YZ1} and \eqref{YZ2} result in \cite{BB}
\begin{equation}\label{gammas}
\gamma_{\hat n}=2\sqrt{(n_{1}+...+n_{d}-1)(n_{1}+...+n_{d})}
\end{equation}
and
\begin{equation}\label{YZ3}
Y_{\hat n}=Z_{\hat n}\sqrt{\frac{n_{1}+...+n_{d}}{n_{1}+...+n_{d}-1}}.
\end{equation}

Let us take a single mode with $\gamma_{\hat n}\neq 0$. Now let us consider formula \eqref{pnumber}, but {\em without} the nonlinear corrections (from here and below, the corresponding expressions will be denoted as ${N'}_{p}^{\,\hat n}$ instead of $N_{p}^{\hat n}$). For a single mode one gets
\begin{equation}\label{partnum-lin}
{N'}_{p}^{\,\hat n}=\frac{\alpha^{2}}{2}\int\left(\xi_{\hat n}\xi_{\hat n}^{*}+\eta_{\hat n}\eta_{\hat n}^{*}\right)d^{d}x+
\frac{\alpha^{2}}{4}\left[e^{-2i\gamma_{\hat n}t}\int(\xi_{\hat n}^{2}-\eta_{\hat n}^{2})d^{d}x+\textrm{c.c.}\right].
\end{equation}
Using \eqref{YZG} and \eqref{YZ3}, we arrive at
\begin{equation*}
\int(\xi_{\hat n}^{2}-\eta_{\hat n}^{2})d^{d}x=\frac{Z_{\hat n}^{2}}{n_{1}+...+n_{d}-1}\int G_{\hat n}^{2}(\vec x)d^{d}x\neq 0.
\end{equation*}
Thus, the time-dependent terms in \eqref{partnum-lin} do not vanish. Consequently, without the nonlinear corrections energy \eqref{energyprelim} of even a single mode also is not conserved over time. This example explicitly demonstrates that in the general case the integrals of motion calculated up to the quadratic order in perturbations indeed are not conserved over time if only the linear part of the perturbation is taken into account.

Now let us calculate $N_{p}$, $E_{p}$ and $P_{p,l}$ using the correct formulas \eqref{PNfinal}, \eqref{Efinal} and \eqref{Pfinal}. According to \eqref{PNfinal} with \eqref{log-pot-F} and \eqref{sol-pot-F},
\begin{equation}\label{Npzero}
N_{p}^{\,\hat n}=0
\end{equation}
for any mode, leading to $N_{p}=0$. Note that \eqref{Npzero} is not a general rule, it is just a curious property of the model at hand. Analogously, from \eqref{Pfinal} with \eqref{YZG} and \eqref{YZ3}, one can get
\begin{equation*}
P_{p,l}^{\,\hat n}=0
\end{equation*}
for any mode (this result is expected, because all modes are localized on the soliton), leading to $P_{p,l}=0$. And finally, from \eqref{Efinal} with \eqref{YZG}, \eqref{gammas} and \eqref{YZ3}, one can get
\begin{equation*}
E_{p}=2\alpha^{2}\sum_{\hat n}(n_{1}+...+n_{d})Z_{\hat n}^{*}Z_{\hat n}\int G_{\hat n}^{2}(\vec x)d^{d}x.
\end{equation*}

\subsection{The Gross--Pitaevskii equation}
Now we turn to a much more physically motivated example. Let us consider the Gross--Pitaevskii equation
\begin{equation*}
i\frac{\partial\Psi}{\partial t}=-\Delta\Psi+g|\Psi|^{2}\Psi,
\end{equation*}
where  $d=3$ and $g>0$. For a spatially uniform case, the stationary solution takes the standard form
\begin{equation}\label{GPsol}
\Psi_{0}(t,\vec x)=e^{-i\omega t}\sqrt{\frac{\omega}{g}}
\end{equation}
with $\omega>0$. For simplicity, let us suppose that the volume of the system is finite. The particle number and the energy, corresponding to solution \eqref{GPsol}, look as follows:
\begin{equation}
N_{0}=\frac{\omega}{g}L^{3},\qquad E_{0}=\frac{\omega^{2}}{2g}L^{3}=\frac{g}{2L^{3}}N_{0}^{2},
\end{equation}
where $L^{3}$ is the volume of the system.

Let us take the linear part of the perturbation in the form of a superposition of the plane waves \cite{Bogolyubov}
\begin{equation}\label{pertGP}
\varphi_{lin}(t,\vec x)=\frac{1}{\sqrt{L^{3}}}\sum\limits_{j}\left(a_{j}e^{-i(\gamma_{j}t-\vec k_{j}\vec x)}+b_{j}e^{i(\gamma_{j}t-\vec k_{j}\vec x)}\right)
\end{equation}
with periodic boundary conditions, where $a_{j}$ and $b_{j}$ are complex coefficients. Linearized equations of motion \eqref{eqxin} and \eqref{eqetan} for perturbation \eqref{pertGP} reduce to the system of equations
\begin{align}\label{eqab1}
&({\vec k_{j}}^{2}+\omega-\gamma_{j})a_{j}+\omega b_{j}^{*}=0,\\\label{eqab2}
&\omega a_{j}+({\vec k_{j}}^{2}+\omega+\gamma_{j})b_{j}^{*}=0,.
\end{align}
As expected, this system of equations leads to the famous Bogolyubov dispersion law \cite{Bogolyubov}
\begin{equation}\label{DispLaw}
\gamma_{j}=\sqrt{{\vec k_{j}}^{2}(2\omega+{\vec k_{j}}^{2})}.
\end{equation}
For $\vec k_{0}=0$, we get $a_{0}+b_{0}^{*}=0$. This solution describes the mode $if$, corresponding to the global $U(1)$ symmetry of the theory.

Using equations \eqref{PNfinal}, \eqref{Efinal}, \eqref{Pfinal} and \eqref{eqab1}--\eqref{DispLaw}, after some algebra we can obtain (in what follows, for simplicity I will omit the small parameter $\alpha$ that was used previously)
\begin{align}\label{Np}
&N_{p}^{j}=-(a_{j}+b_{j}^{*})(a_{j}^{*}+b_{j})=-\frac{|\vec k_{j}|}{\sqrt{2\omega+{\vec k_{j}}^{2}}}\,\tilde n_{j}\leq 0,\\\label{Ep}
&E_{p}^{j}=\omega N_{p}^{j}+\gamma_{j}(a_{j}^{*}a_{j}-b_{j}^{*}b_{j})=\omega N_{p}^{j}+\gamma_{j}\tilde n_{j},\\\label{Pp}
&\vec P_{p}^{j}=\vec k_{j}(a_{j}^{*}a_{j}-b_{j}^{*}b_{j})=\vec k_{j}\tilde n_{j},
\end{align}
where $a_{j}^{*}a_{j}-b_{j}^{*}b_{j}=\tilde n_{j}>0$ for $\vec k_{j}\neq 0$. In quantum theory, $\tilde n_{j}$ corresponds to the number of quasi-particles with the momentum $\vec k_{j}$. The energy of a single mode $E_{p}^{j}$ can be rewritten as
\begin{equation}\label{Ep2}
E_{p}^{j}=\gamma_{j}\tilde n_{j}\frac{\omega+{\vec k_{j}}^{2}}{2\omega+{\vec k_{j}}^{2}}=\sqrt{\frac{\omega}{2}}\,\bigl|\vec P_{p}^{j}\bigr|\frac{1+\frac{{\vec k_{j}}^{2}}{\omega}}{\sqrt{1+\frac{{\vec k_{j}}^{2}}{2\omega}}}\ge\sqrt{\frac{\omega}{2}}\,\bigl|\vec P_{p}^{j}\bigr|\ge 0.
\end{equation}
For small $\vec k_{j}$, the mode has the phonon-like spectrum.

Thus, for the total energy of the system $E=E_{0}+\sum_{j}{E}_{p}^{j}$ and its total momentum $\vec P$ we can write with the same accuracy (because $|N_{p}|=\sum_{j}|{N}_{p}^{j}|\ll N_{0}$)
\begin{align}\label{Etot}
&E=\frac{g}{2L^{3}}N^{2}+\sum\limits_{j}\gamma_{j}\tilde n_{j},\\\label{Ptot}
&\vec P=\sum\limits_{j}\vec k_{j}\tilde n_{j},
\end{align}
where $N=N_{0}+N_{p}<N_{0}$. As in the quantum theory \cite{Bogolyubov}, here the term $\omega N_{p}$ is formally incorporated into the energy of the ground state $\frac{g}{2L^{3}}N^{2}$. According to formula \eqref{Etotfinal}, the difference between the total energy of the final configuration and the energy of the initial configuration in the processes which conserve the total particle number is just $\sum_{j}\gamma_{j}\tilde n_{j}$.

As was noted at the end of the previous section, some of the time-dependent terms in formulas \eqref{pnumber} and \eqref{momentum} may vanish even without the nonlinear corrections $\chi_{\pm,n}$, $\psi_{\pm,n}$, $\varrho_{\pm,nk}$ and $\theta_{\pm,nk}$ if a perturbation consists of just the plane waves. For example, for a perturbation which contains a single linear mode from \eqref{pertGP}, instead of the particle number of the mode $N_{p}^{j}$ defined by \eqref{Np} we obtain
\begin{equation}\label{Npconsincorr}
{N'}_{p}^{\,j}=a_{j}^{*}a_{j}+b_{j}^{*}b_{j}>0.
\end{equation}
In particular, for $0<|\vec k_{j}|\ll \sqrt{2\omega}$ we get
\begin{equation*}
{N'}_{p}^{\,j}\simeq \sqrt{\frac{\omega}{2}}\,\tilde n_{j}\frac{1}{|\vec k_{j}|}.
\end{equation*}
For $\vec k_{0}=0$, we obtain
\begin{equation*}
{N'}_{p}^{\,0}=2a_{0}^{*}a_{0}>0,
\end{equation*}
which looks unphysical, because different solutions which are symmetric with respect to the global $U(1)$ symmetry should have the same values of the integrals of motion. Meanwhile, from the correct formula \eqref{Np} we get
\begin{equation}
N_{p}^{j}\simeq-\frac{1}{\sqrt{2\omega}}\,\tilde n_{j}|\vec k_{j}|
\end{equation}
for $|\vec k_{j}|\ll \sqrt{2\omega}$ and the expected result $N_{p}^{0}=0$ for $\vec k_{0}=0$.

\section{Conclusion}
In the present paper, perturbations against a stationary solution of the nonlinear Schr\"{o}dinger equation with the general form of nonlinearity are examined. It is shown that the use of nonlinear equations of motion for the perturbations is necessary for obtaining correct and conserved over time nonzero expressions for the integrals of motion even in the quadratic order in the corresponding expansion parameter, as well as for the validity of the additivity property for these integrals of motion. As noted in the Introduction, this effect may indicate some sort of the nonlinear superposition principle. It could be also important for a correct quantization of perturbations against stationary solutions of the nonlinear Schr\"{o}dinger equation.

As a demonstration of these results, two explicit examples, --- the case of logarithmic nonlinearity and the Gross--Pitaevskii equation, are considered. One can see that the use of nonlinear equations of motion for perturbations indeed recovers the conservation over time of the particle number even in the case of perturbation consisting of only a single nonlinear mode, compare \eqref{partnum-lin} with \eqref{Npzero}. When the use of only the linear approximation provides a result that is conserved over time, like the one in \eqref{Npconsincorr}, the correct result may look completely different (compare the signs of \eqref{Np} and \eqref{Npconsincorr}). Note that the nonlinear correction is of the same order as the ``main'' result and can even fully compensate it (as in \eqref{Npzero}).

Finally, one may assume that in some cases the additivity property can survive at higher orders of perturbation theory: the overlap integrals between different modes may appear only together with the corresponding time-dependent exponential factors, which should vanish because the particle number, energy and momentum are conserved over time. This question calls for further investigation.

\section*{Acknowledgements}
The author is grateful to D.G.~Levkov and I.P.~Volobuev for useful comments. The work was supported by the Grant 16-12-10494 of the Russian Science Foundation.

\section*{Appendix~A}
\begin{enumerate}
\item
Let us take equation \eqref{psiminuseq}, multiply it by $f$ and integrate the result over the space. Using the fact that $\hat L_{2}f=0$, we get
\begin{equation}\label{appA1}
2\gamma_{n}\int f\psi_{+,n}d^{d}x=\int fW\xi_{n}\eta_{n}d^{d}x.
\end{equation}
Then, using equations \eqref{eqxin} and \eqref{eqetan}, we get
\begin{equation}\label{appA2}
\int\left(\eta_{n}\hat L_{1}\xi_{n}-\xi_{n}\hat L_{2}\eta_{n}\right)d^{d}x=\gamma_{n}\int\left(\eta_{n}^{2}-\xi_{n}^{2}\right)d^{d}x.
\end{equation}
From definition \eqref{Leq} it follows that
\begin{equation}\label{appBLdefeq}
\int\left(\eta_{n}\hat L_{1}\xi_{n}-\xi_{n}\hat L_{2}\eta_{n}\right)d^{d}x=2\int fW\xi_{n}\eta_{n}\,d^{d}x.
\end{equation}
Finally, combining relations \eqref{appA1}--\eqref{appBLdefeq} and using the fact that $\gamma_{n}\neq 0$, we obtain
\begin{equation*}
\int \left(f\psi_{+,n}+\frac{1}{4}\left(\xi_{n}^{2}-\eta_{n}^{2}\right)\right)d^{d}x=0.
\end{equation*}

\item
Let us take equation \eqref{eqvarrho}, multiply it by $f$ and integrate the result over the space. Using the fact that $\hat L_{2}f=0$, we get
\begin{equation}\label{auxil0}
(\gamma_{n}+\gamma_{k})\int f\varrho_{+,nk}\,d^{d}x=
\int fW\left(\xi_{n}\eta_{k}+\eta_{n}\xi_{k}\right)d^{d}x.
\end{equation}
Then, using equations \eqref{eqxin}, \eqref{eqetan}, we get
\begin{equation*}
\int\left(\eta_{k}\hat L_{1}\xi_{n}-\xi_{n}\hat L_{2}\eta_{k}\right)d^{d}x=\int(\gamma_{n}\eta_{k}\eta_{n}-\gamma_{k}\xi_{n}\xi_{k})d^{d}x,
\end{equation*}
From definition \eqref{Leq} it follows that
\begin{equation*}
\int\left(\eta_{k}\hat L_{1}\xi_{n}-\xi_{n}\hat L_{2}\eta_{k}\right)d^{d}x=2\int fW\xi_{n}\eta_{k}\,d^{d}x,
\end{equation*}
leading to
\begin{equation}\label{auxil1}
2\int fW\xi_{n}\eta_{k}\,d^{d}x=\int \left(\gamma_{n}\eta_{n}\eta_{k}-\gamma_{k}\xi_{n}\xi_{k}\right)d^{d}x.
\end{equation}
Analogously, we get
\begin{equation}\label{auxil2}
2\int fW\xi_{k}\eta_{n}\,d^{d}x=\int \left(\gamma_{k}\eta_{n}\eta_{k}-\gamma_{n}\xi_{n}\xi_{k}\right)d^{d}x.
\end{equation}
Summing up relations \eqref{auxil1} and \eqref{auxil2}, we obtain
\begin{equation}\label{auxil3}
\int fW(\xi_{n}\eta_{k}+\xi_{k}\eta_{n})d^{d}x=\frac{\gamma_{n}+\gamma_{k}}{2}\int\left(\eta_{n}\eta_{k}-\xi_{n}\xi_{k}\right)d^{d}x.
\end{equation}
Finally, combining relations \eqref{auxil0}, \eqref{auxil3} and using the fact that $\gamma_{n}+\gamma_{k}\neq 0$, we arrive at
\begin{equation*}
\int\left(f\varrho_{+,nk}+\frac{1}{2}\left(\xi_{n}\xi_{k}-\eta_{n}\eta_{k}\right)\right)d^{d}x=0.
\end{equation*}

\item
Let us take equation \eqref{eqtheta}, multiply it by $f$ and integrate the result over the space. Using the fact that $\hat L_{2}f=0$, we get
\begin{equation}\label{auxil4}
(\gamma_{n}-\gamma_{k})\int f\theta_{+,nk}\,d^{d}x=\int fW\left(\eta_{n}\xi_{k}^{*}-\xi_{n}\eta_{k}^{*}\right)d^{d}x.
\end{equation}
Then, using equations \eqref{eqxin}, \eqref{eqetan}, we get
\begin{equation*}
\int\left(\eta_{n}\hat L_{1}\xi_{k}^{*}-\xi_{k}^{*}\hat L_{2}\eta_{n}\right)d^{d}x=\int(\gamma_{k}\eta_{n}\eta_{k}^{*}-\gamma_{n}\xi_{k}^{*}\xi_{n})d^{d}x.
\end{equation*}
From definition \eqref{Leq} it follows that
\begin{equation*}
\int\left(\eta_{n}\hat L_{1}\xi_{k}^{*}-\xi_{k}^{*}\hat L_{2}\eta_{n}\right)d^{d}x=2\int fW\xi_{k}^{*}\eta_{n}\,d^{d}x.
\end{equation*}
leading to
\begin{equation}\label{auxil5}
2\int fW\xi_{k}^{*}\eta_{n}\,d^{d}x=\int \left(\gamma_{k}\eta_{n}\eta_{k}^{*}-\gamma_{n}\xi_{n}\xi_{k}^{*}\right)d^{d}x.
\end{equation}
Analogously, we get
\begin{equation}\label{auxil6}
2\int fW\xi_{n}\eta_{k}^{*}\,d^{d}x=\int \left(\gamma_{n}\eta_{n}\eta_{k}^{*}-\gamma_{k}\xi_{n}\xi_{k}^{*}\right)d^{d}x.
\end{equation}
Subtracting \eqref{auxil6} from \eqref{auxil5}, we obtain
\begin{equation}\label{auxil9}
\int fW(\xi_{k}^{*}\eta_{n}-\xi_{n}\eta_{k}^{*})d^{d}x=\frac{\gamma_{k}-\gamma_{n}}{2}\int \left(\eta_{n}\eta_{k}^{*}+\xi_{n}\xi_{k}^{*}\right)d^{d}x.
\end{equation}
Finally, combining relations \eqref{auxil4}, \eqref{auxil9} and using the fact that $\gamma_{n}-\gamma_{k}\neq 0$, we arrive at
\begin{equation*}
\int\left(f\theta_{+,nk}+\frac{1}{2}(\xi_{n}\xi_{k}^{*}+\eta_{n}\eta_{k}^{*})\right) d^{d}x=0.
\end{equation*}
\end{enumerate}

\section*{Appendix~B}
\begin{enumerate}
\item
Using equation \eqref{eqxin}, one can get
\begin{equation*}
\int \gamma_{k}\xi_{n}\eta_{k}\,d^{d}x=\int\xi_{k}\hat L_{1}\xi_{n}d^{d}x=\int \gamma_{n}\xi_{k}\eta_{n}\,d^{d}x,
\end{equation*}
leading to
\begin{equation}\label{auxen1}
\int\left(\gamma_{n}\eta_{n}\xi_{k}-\gamma_{k}\eta_{k}\xi_{n}\right)d^{d}x=0.
\end{equation}
Analogously, using equation \eqref{eqetan}, one can get
\begin{equation}\label{auxen2}
\int\left(\gamma_{n}\eta_{k}\xi_{n}-\gamma_{k}\eta_{n}\xi_{k}\right)d^{d}x=0.
\end{equation}
By subtracting \eqref{auxen2} from \eqref{auxen1} and using the fact that $\gamma_{n}+\gamma_{k}\neq 0$, we obtain
\begin{equation*}
\int\left(\eta_{n}\xi_{k}-\eta_{k}\xi_{n}\right)d^{d}x=0.
\end{equation*}
Actually, relations \eqref{auxen1} and \eqref{auxen2} lead to a more stringent condition
\begin{equation*}
\int\eta_{n}\xi_{k}d^{d}x=0
\end{equation*}
for $\gamma_{n}\neq\gamma_{k}$.

\item
Analogously, we can get
\begin{equation}\label{auxen3}
\int\left(\gamma_{n}\eta_{n}\xi_{k}^{*}-\gamma_{k}\eta_{k}^{*}\xi_{n}\right)d^{d}x=0
\end{equation}
and
\begin{equation}\label{auxen4}
\int\left(\gamma_{n}\eta_{k}^{*}\xi_{n}-\gamma_{k}\eta_{n}\xi_{k}^{*}\right)d^{d}x=0.
\end{equation}
By adding \eqref{auxen3} to \eqref{auxen4} and using the fact that $\gamma_{n}-\gamma_{k}\neq 0$, we obtain
\begin{equation*}
\int\left(\eta_{n}\xi_{k}^{*}+\eta_{k}^{*}\xi_{n}\right)d^{d}x=0.
\end{equation*}
\end{enumerate}

\section*{Appendix~C}
All the calculations presented below are valid only for $V(\vec x)\equiv 0$.

First, we take equation \eqref{eqxin} and differentiate it with respect to $x^{l}$. We get
\begin{equation}\label{L1diffi}
\hat L_{1}\partial_{l}\xi_{n}+2\partial_{l}f\left(3W+4J\right)\xi_{n}=\gamma_{n}\partial_{l}\eta_{n}.
\end{equation}
Second, we take equation \eqref{eqetan} and differentiate it with respect to $x^{l}$. We get
\begin{equation}\label{L2diffi}
\hat L_{2}\partial_{l}\eta_{n}+2\partial_{l}fW\eta_{n}=\gamma_{n}\partial_{l}\xi_{n}.
\end{equation}
\begin{enumerate}
\item
Let us take equation \eqref{psipluseq}, multiply it by $\partial_{l}f$ and integrate the result over the space. Using the fact that $\hat L_{1}\partial_{l}f=0$, we get
\begin{equation}\label{appD1}
2\gamma_{n}\int\partial_{l}f\psi_{-,n}d^{d}x=\int\partial_{l}f\left(W\left(\frac{3}{2}\xi_{n}^{2}-\frac{1}{2}\eta_{n}^{2}\right)+2J\xi_{n}^{2}\right)d^{d}x.
\end{equation}
Now we take equation \eqref{L1diffi}, multiply it by $\xi_{n}$, integrate the result over the space and apply equation \eqref{eqxin}. We get
\begin{equation}\label{appD2}
\gamma_{n}\int\xi_{n}\partial_{l}\eta_{n}d^{d}x=\int\partial_{l}f\left(3W\xi_{n}^{2}+4J\xi_{n}^{2}\right)d^{d}x.
\end{equation}
Then we take equation \eqref{L2diffi}, multiply it by $\eta_{n}$, integrate the result over the space and apply equation \eqref{eqetan}. We get
\begin{equation}\label{appD3}
\gamma_{n}\int\xi_{n}\partial_{l}\eta_{n}d^{d}x=-\int\partial_{l}fW\eta_{n}^{2}\,d^{d}x.
\end{equation}
Substituting relations \eqref{appD2} and \eqref{appD3} into \eqref{appD1} and using the fact that $\gamma_{n}\neq 0$, we arrive at
\begin{equation*}
\int\left(\partial_{l}f\psi_{-,n}-\frac{1}{2}\xi_{n}\partial_{l}\eta_{n}\right)d^{d}x=0.
\end{equation*}

\item
Let us take equation \eqref{eqvarrhoplus}, multiply it by $\partial_{l}f$ and integrate the result over the space. Using the fact that $\hat L_{1}\partial_{l}f=0$, we get
\begin{equation}\label{appD4}
(\gamma_{n}+\gamma_{k})\int\partial_{l}f\varrho_{-,nk}d^{d}x=\int\partial_{l}f\left(W\left(3\xi_{n}\xi_{k}-\eta_{n}\eta_{k}\right)+4J\xi_{n}\xi_{k}\right)d^{d}x.
\end{equation}
Now we take equation \eqref{L1diffi}, multiply it by $\xi_{k}$, integrate the result over the space and apply equation \eqref{eqxin}. We get
\begin{equation}\label{appD5}
\gamma_{n}\int\xi_{k}\partial_{l}\eta_{n}d^{d}x+\gamma_{k}\int\xi_{n}\partial_{l}\eta_{k}d^{d}x=
2\int\partial_{l}f\left(3W\xi_{n}\xi_{k}+4J\xi_{n}\xi_{k}\right)d^{d}x.
\end{equation}
Then we take equation \eqref{L2diffi}, multiply it by $\eta_{k}$, integrate the result over the space and apply equation \eqref{eqetan}. We get
\begin{equation}\label{appD6}
\gamma_{k}\int\xi_{k}\partial_{l}\eta_{n}d^{d}x+\gamma_{n}\int\xi_{n}\partial_{l}\eta_{k}d^{d}x=-2\int\partial_{l}fW\eta_{n}\eta_{k}\,d^{d}x.
\end{equation}
Substituting relations \eqref{appD5} and \eqref{appD6} into \eqref{appD4} and using the fact that $\gamma_{n}+\gamma_{k}\neq 0$, we arrive at
\begin{equation*}
\int\left(\partial_{l}f\varrho_{-,nk}-\frac{1}{2}\left(\xi_{k}\partial_{l}\eta_{n}+\xi_{n}\partial_{l}\eta_{k}\right)\right)d^{d}x=0.
\end{equation*}

\item
Let us take equation \eqref{eqthetaplus}, multiply it by $\partial_{l}f$ and integrate the result over the space. Using the fact that $\hat L_{1}\partial_{l}f=0$, we get
\begin{equation}\label{appD7}
(\gamma_{n}-\gamma_{k})\int\partial_{l}f\theta_{-,nk}d^{d}x=\int\partial_{l}f\left(W\left(3\xi_{n}\xi_{k}^{*}+\eta_{n}\eta_{k}^{*}\right)
+4J\xi_{n}\xi_{k}^{*}\right)d^{d}x.
\end{equation}
Now we take equation \eqref{L1diffi}, multiply it by $\xi_{k}^{*}$, integrate the result over the space and apply the complex conjugate of equation \eqref{eqxin}. We get
\begin{equation}\label{appD8}
\gamma_{n}\int\xi_{k}^{*}\partial_{l}\eta_{n}d^{d}x+\gamma_{k}\int\xi_{n}\partial_{l}\eta_{k}^{*}d^{d}x=
2\int\partial_{l}f\left(3W\xi_{n}\xi_{k}^{*}+4J\xi_{n}\xi_{k}^{*}\right)d^{d}x.
\end{equation}
Then we take equation \eqref{L2diffi}, multiply it by $\eta_{k}^{*}$, integrate the result over the space and apply the complex conjugate of equation \eqref{eqetan}. We get
\begin{equation}\label{appD9}
\gamma_{k}\int\xi_{k}^{*}\partial_{l}\eta_{n}d^{d}x+\gamma_{n}\int\xi_{n}\partial_{l}\eta_{k}^{*}d^{d}x=-2\int\partial_{l}fW\eta_{n}\eta_{k}^{*}\,d^{d}x.
\end{equation}
Substituting relations \eqref{appD8} and \eqref{appD9} into \eqref{appD7} and using the fact that $\gamma_{n}-\gamma_{k}\neq 0$, we arrive at
\begin{equation*}
\int\left(\partial_{l}f\theta_{-,nk}-\frac{1}{2}\left(\xi_{k}^{*}\partial_{l}\eta_{n}-\xi_{n}\partial_{l}\eta_{k}^{*}\right)\right)d^{d}x=0.
\end{equation*}
\end{enumerate}

\end{document}